\documentclass[a4paper, 11pt]{article}
\usepackage{amsmath, amssymb, amsthm}
\usepackage{graphicx}
\usepackage{latexsym}

\def\b{\begin{eqnarray}}
\def\e{\end{eqnarray}}
\def\n{\noindent}
\numberwithin{equation}{section}

\newtheorem{theorem}{Theorem}

\begin{document}

\begin{center}

{\LARGE\textbf{Hamiltonian formulation, nonintegrability and local
bifurcations for the Ostrovsky equation}}

\vspace {10mm} \noindent {\large \bf Roy Choudhury, $^{\dagger}$ Rossen I. Ivanov,
$^{\dagger \dagger}$\footnote{On leave from the Institute for Nuclear
Research and Nuclear Energy, Bulgarian Academy of Sciences, Sofia,
Bulgaria.} Yue Liu $^{\ast}$}

\begin{tabular}{c}
\\$\phantom{R^R}^\dagger ${Department of Mathematics, University of Central Florida, Orlando, FL 32816} \\ {\it choudhur@longwood.cs.ucf.edu}
\\
$\phantom{R^R}^{\dagger \dagger} ${School of Mathematics, Trinity
College Dublin, Dublin 2
Ireland} \\ {\it ivanovr@tcd.ie} \\
$\phantom{R^R}^\ast$Department of Mathematics, University of Texas, Arlington, TX 76019 \\ {\it yliu@uta.edu} \\
\end{tabular}
\vskip1cm
\end{center}

\vskip1cm

\begin{abstract}
\noindent The Ostrovsky equation is a model for gravity waves
propagating down a channel under the influence of Coriolis force.
This equation is a modification of the famous Korteweg-de Vries
equation and is also Hamiltonian. However the Ostrovsky equation is
not integrable and in this contribution we prove its
nonintegrability. We also study local bifurcations of its solitary
waves.

{\bf MSC:} 35Q35, 35Q53, 37K10

{\bf Key Words:} Conservation Laws, Integrability, Local bifurcation, Ostrovsky
equation.

\end{abstract}

\section{Introduction}

In this contribution we investigate the integrability and the local
bifurcations of the solitary waves of the nonlinear equation
\begin{equation}\label{eq1}
 (u_{t}-\beta u_{xxx}+2uu_x)_x=\gamma u,
\end{equation}
where $\beta$ and $\gamma$  are constant parameters. The equation
(\ref{eq1}) is known as the Ostrovsky equation \cite{O78}, and is
often called the Rotation-Modified Korteweg-de Vries equation. It is
a model for gravity waves propagating down a channel under the
influence of Coriolis force.  In essence, $u$ in the equation can be
regarded as the fluid velocity in the $x-$direction. The physical
parameter $\gamma >  0$ measures the effect of the Earth's rotation.
The parameter $\beta $ determines the type of dispersion, namely
$\beta < 0 $ (negative-dispersion) for surface and internal waves in
the ocean and surface waves in a shallow channel with an uneven
bottom. The parameter value $\beta > 0 $ (positive dispersion)
applies to capillary waves on the surface of liquid or  oblique
magneto-acoustic waves in plasma. In the limit $\gamma = 0$,
(\ref{eq1}) apparently reduces to the integrable Korteweg-de Vries
equation. More details about the Ostrovsky equation can be found in
\cite{Be, GaSt91, GiGrSt, Gr, GrOsShSt}.

The equation (\ref{eq1}) has three known integrals of motion.
Indeed, it can be formally rewritten in a conservation law form as
follows ($\partial \equiv \partial/\partial x$):

\b\label{eq2a} u_t\!\!\!&=&\!\!\! \Big(\frac{1}{\gamma}(u_t-\beta
u_{xxx}+2 u
u_x)_t \Big)_x , \\
\label{eq2b} \Big(\frac{u^2}{2} \Big)_t \!\!\!&=&\!\!\! \Big(\beta
(u
u_{xx}-\frac{1}{2}u_x^2)-\frac{2}{3}u^3+\frac{\gamma}{2}(\partial^{-1}u)^2
\Big)_x ,\\
\label{eq2c} \Big(\frac{\beta}{2}
u_x^2\!\!\!&+&\!\!\!\frac{\gamma}{2}(\partial^{-1}u)^2 +
\frac{u^3}{3} \Big)_t=\Big[ \beta^2(u_x u_{xxx}-\frac{1}{2}u_{xx}^2)
\phantom{**************}\nonumber
\\\!\!\!&-&\!\!\!\beta(2 u u_x^2 -u^2 u_{xx})+\beta\gamma u_x(\partial^{-1}u)
+\frac{\gamma^2}{2}(\partial^{-2}u)^2-\frac{u^4}{2}\Big]_x . \e

\n From (\ref{eq2a}) -- (\ref{eq2c}) we have the following three
integrals, provided the solution $u$ is in an appropriate class of
functions, such that the integration (over the real line for
Schwartz class functions, or over a period for periodic functions)
makes sense (e.g. see \cite{GL05} for details): \b
\label{eqI}I[u]&=&\int u\text{d}x=0,\qquad \gamma \neq 0,
 \\P[u]&=&\frac{1}{2}\int u^2 \text{d}x , \label{eqP}\\
H[u]&=&\int \Big(\frac{\beta}{2}
u_x^2+\frac{\gamma}{2}(\partial^{-1}u)^2 + \frac{u^3}{3}
\Big)\text{d}x .\label{eqH}\e

Now we notice that the equation (\ref{eq1}) can be written as
\begin{equation}\label{eqH1}
 u_{t}=\frac{\partial}{\partial x}\frac{\delta H}{\delta u(x)},
\end{equation}

\n where the symbol $\delta/\delta u$ denotes variational
derivative. Moreover, (\ref{eqH1}) can be further represented in a
Hamiltonian form with a Hamiltonian $H$:
\begin{equation}\label{eqH2}
 u_{t}=\{u, H\}.
\end{equation}
\n The Poisson bracket is defined as
\begin{equation}\label{eqH3}
 \{F, G\}\equiv \int\frac{\delta F}{\delta u(x)}\frac{\partial}{\partial x}\frac{\delta G}{\delta
 u(x)}\text{d}x.
\end{equation}

\n One can check that the bracket (\ref{eqH3}) is anti-symmetric and
satisfies Jacobi identity.

Also, we notice that the integral (\ref{eqP}) has a meaning of a
momentum. Indeed, it is related to the translation invariance of the
Hamiltonian. Since $H[u(x+\varepsilon)]-H[u(x)]\equiv 0$, the
expansion of $\int(H[u(x+\varepsilon)]-H[u(x)])\text{d}x$  in
$\varepsilon$ about $\varepsilon=0$ gives (note that $u(x)=\delta
P/\delta u$)

\begin{equation}
0=\int\frac{\delta H}{\delta u(x)}\frac{\partial u}{\partial
x}\text{d}x=\int\frac{\delta H}{\delta u(x)} \frac{\partial
}{\partial x}\frac{\delta P}{\delta u(x)}\text{d}x\equiv \{H,
P\}=-P_t,\nonumber
\end{equation}

\n i.e. $P_t=0$. The existence of (\ref{eqI}) from the Hamiltonian
viewpoint is related to the presence of the operator
$\partial/\partial x$ in the Poisson brackets.

Apparently there is no second Hamiltonian formulation for the
Ostrovsky equation, compatible with the presented one, i.e. this
equation is not bi-Hamiltonian. Despite the existence of the three
integrals, in Section 2 we demonstrate that the equation (\ref{eq1})
is not completely integrable for $\gamma\neq 0$. In Section 3 we
study the local bifurcations of its solitary waves.

\section{The Integrability Test}

In our analysis we use the integrability check developed in
\cite{MN02, SJ98, OJ00}. (Another application of the method is
presented in \cite{I05}.)

This perturbative method can be briefly outlined as  follows.
Consider the evolution partial differential equation
\begin{equation}\label{eq3}
u_{t}=F_{1}[u]+F_{2}[u]+F_{3}[u]+\ldots
\end{equation}
where $F_{k}[u]$ is a homogeneous differential polynomial, i.e. a
polynomial of the variables $u$, $u_{x}$, $u_{xx}$, ...,
$\partial^{n}_{x}u$ with complex constant coefficients, satisfying
the condition
\begin{equation}
F_{k}[\lambda u]=\lambda ^{k}F_{k}[u], \qquad \lambda \in
\mathbb{C}. \nonumber
\end{equation}
The linear part is $F_{1}[u]=L(u)$, where $L$ is a linear
differential operator of order two or higher. The representation
(\ref{eq3}) can be put into correspondence to a symbolic expression
of the form

\begin{equation}\label{eq5}
 u_{t}=u\omega (\xi_{1})+\frac{u^{2}}{2}a_{1}(\xi _{1}, \xi _{2})+ \frac{u^{3}}{3}a_{2}(\xi _{1}, \xi _{2}, \xi_{3})+\ldots=F
\end{equation}
where $\omega (\xi_{1})$ is a polynomial of degree two or higher and
$a_{k}(\xi _{1}, \xi _{2},\ldots \xi _{k+1})$ are symmetric
polynomials. Each of these polynomials is related to the Fourier
image of the corresponding $F_{k}[u]$ and can be obtained through a
simple procedure, described e.g. in \cite{MN02}. Each differential
monomial $u^{n_{0}}u_{x}^{n_{1}}\ldots(\partial_{x}^{q}u)^{n_{q}}$
is represented by a symbol
\begin{equation}
u^{m}\langle \xi_{1}^{0}\ldots
\xi_{n_{0}}^{0}\xi_{n_{0}+1}^{1}\ldots
\xi_{n_{0}+n_{1}}^{1}\xi_{n_{0}+n_{1}+1}^{2}\ldots
\xi_{n_{0}+n_{1}+n_{2}}^{2}\ldots \xi_{m}^{q}\rangle \nonumber
\end{equation}
where $m=n_{0}+n_{1}+\ldots +n_{q}$ and the brackets $\langle  \
\rangle$ denote symmetrization over all arguments $\xi_{k}$ (i.e.
symmetrization with respect to the group of permutations of $m$
elements $S_{m}$):
\begin{equation}
\langle f(\xi_{1},\xi_{2},\ldots,
\xi_{n})\rangle=\frac{1}{m!}\sum_{\sigma\in S_{m}}
f(\xi_{\sigma(1)},\xi_{\sigma(2)},\ldots, \xi_{\sigma(n)}) \nonumber
\end{equation}
Also, for any function $F$ (\ref{eq5}) there exists a formal
recursion operator
\begin{equation}\label{eq6}
 \Lambda = \eta +u \phi_{1}(\xi_{1},\eta)+u^{2}\phi_{2}(\xi _{1}, \xi _{2},\eta)+ \ldots
\end{equation}
where the coefficients $\phi_{m}(\xi _{1}, \xi _{2},\ldots \xi _{m},
\eta)$ can be determined recursively:

\begin{subequations} \label{eq7}
\begin{gather}
  \phi_{1}(\xi_{1},\eta) = N^{\omega}(\xi_{1},\eta)\xi_{1}a_{1}(\xi_{1},\eta)      \label{eq6a}\\
 \phi_{m}(\xi _{1}, \xi _{2},\ldots \xi _{m},\eta) = \phantom{*************************************}\nonumber \\
 N^{\omega}(\xi _{1}, \xi _{2},\ldots \xi _{m},\eta) \Big \{ (\xi _{1}+ \xi _{2}+\ldots +\xi _{m})a_{m}(\xi _{1}, \xi _{2},\ldots \xi
 _{m},\eta) + \nonumber \\
 +\sum_{n=1}^{m-1}\Big \langle \frac{n}{m-n+1}\phi_{n}(\xi _{1},\ldots \xi _{n-1},\xi _{n}+\ldots+\xi _{m},\eta)
  a_{m-n}(\xi _{n},\ldots \xi _{m})  + \nonumber \\
+\phi_{n}(\xi _{1},\ldots \xi _{n},\eta+\xi _{n+1}+\ldots+\xi_{m})
   a_{m-n}(\xi _{n+1},\ldots \xi _{m},\eta)  - \nonumber \\
  -\phi_{n}(\xi _{1},\ldots \xi _{n},\eta)
   a_{m-n}(\xi _{n+1},\ldots \xi _{m},\eta+\xi _{1}+\ldots+\xi_{n})\Big \rangle\Big\}    \label{eq6c}
\end{gather}
\end{subequations}
with
\begin{equation}\label{eq8}
N^{\omega}(\xi _{1}, \xi _{2},\ldots \xi _{m}) =
\Big(\omega(\sum_{n=1}^{m}\xi _{n})-\sum_{n=1}^{m}\omega
(\xi_{n})\Big)^{-1}
\end{equation}
and the symbols $\langle \ \ \rangle$ denote symmetrization with
respect to $\xi _{1}, \xi _{2},\ldots \xi _{m}$, (the symbol $\eta$
is not included in the symmetrization).

We  need also the notion of a {\it local} function, which can be
defined as follows. The function $b_{m}(\xi _{1}, \xi _{2},\ldots
\xi _{m},\eta)$ , $m\geq 1$ is called local if all coefficients
$b_{mn}(\xi_{1},\xi_{2},\ldots \xi _{m})$, $n=n_{s},n_{s+1},\ldots$
of its expansion as $\eta\rightarrow \infty$
\begin{equation}\label{eq9}
 b_{m}(\xi _{1}, \xi _{2},\ldots \xi _{m},\eta) =\sum_{n=n_{s}}^{\infty}b_{mn}(\xi _{1}, \xi _{2},\ldots \xi _{m})\eta^{-n}
\end{equation}
are symmetric polynomials.

The integrability criterion can be summarized as follows
\cite{MN02}:
\begin{theorem} \label{th1}
 The complete integrability of the equation (\ref{eq3}), i.e. the existence of
an infinite hierarchy of local symmetries or conservation laws,
implies that all the coefficients (\ref{eq7}) of the formal
recursion operator(\ref{eq6}) are local.
\end{theorem}

The equation (\ref{eq1}) can be written in the form
\begin{equation}\label{eq10}
u_{t}=\partial ^{-1}(\beta \partial ^4 +\gamma) u- 2uu_x
\end{equation}
The symbolic representation of the operator $\partial^{-1}$ is
$1/\eta $. Moreover, Theorem \ref{th1} can be applied in this case
as well \cite{MN02}. The equation (\ref{eq10}) can be represented in
the form (\ref{eq5}) with
\begin{subequations} \label{eq12}
\begin{gather}
\omega(\xi_{1})=\frac{\beta \xi_{1}^4+\gamma }{\xi_{1}}      \label{eq12a}\\
a_{1}(\xi _{1}, \xi _{2})=-2(\xi _{1}+ \xi _{2})      \label{eq12b}
\end{gather}
\end{subequations}
Then from (\ref{eq7}):
\begin{subequations} \label{eq13}
\begin{gather}
 \phi_{1}(\xi_{1},\eta)=-\frac{2\xi_{1}^{2}\eta(\xi_{1}+\eta)^2}
{3\beta \eta^2 \xi_{1}^{2}(\xi_{1}+\eta)^2-\beta \gamma(\xi_{1}+\eta)^2+\gamma \eta\xi_1}      \label{eq13a}\\
  \phi_{2}(\xi_{1},\xi_{2},\eta)=\Phi_{2,-3}(\xi_{1},\xi_{2})\eta^{-3} + \Phi_{2,-4}(\xi_{1},\xi_{2})\eta ^{-4}+\ldots \label{eq13b}
 \end{gather}
\end{subequations}

The expansion of $\phi_{1}(\xi_{1},\eta)$ (\ref{eq13a}) with respect
to $\eta$ is
\begin{equation}\label{eq14}
\phi_{1}(\xi_{1},\eta)=-\frac{2}{3\beta}\eta^{-1}-\frac{2\gamma}{9\beta^2\xi_1^2}\eta^{-3}+\frac{2\gamma}{9\beta^2\xi_1}\eta^{-4}+\frac{2\gamma(\gamma+6\beta\xi_1^4)}{27\beta^3\xi_1
^4}\eta^{-5}+\ldots
\end{equation}
\n and therefore there are obstacles to the integrability of
(\ref{eq1}), since the coefficients in this expansion are not
polynomials in $\xi_1$ for $\gamma \neq 0$.

For comparison, in the integrable case $\gamma=0$ (KdV), we have
\begin{subequations} \label{eq15}
\begin{gather}
 \phi_{1}(\xi_{1},\eta)=-\frac{2} {3\beta} \eta^{-1}  \label{eq15a}\\
\phi_{2}(\xi_{1},\xi_{2},\eta)=-\frac{4} {9\beta^2}\eta^{-3} + \frac{4}{9\beta^2}(\xi_{1}+\xi_{2})\eta^{-4}-\frac{4}{9\beta^2}(\xi_{1}^2+\xi_1\xi_1+\xi_{2}^2)\eta ^{-5}+\nonumber \\
 + \frac{4}{9\beta^2}(\xi_{1}+\xi_{2})(\xi_{1}^2+\xi_{2}^2)\eta ^{-6}+\ldots \label{eq15b}
\end{gather}
\end{subequations}

Therefore, the only completely integrable equation of the form
(\ref{eq1}) is the one with $\gamma=0$, i.e. the KdV equation.

\section{Solitary Waves and Local Bifurcations}

Solitary waves of the Ostrovsky equation of the form $ u(x, t) =
\phi(x-ct) \equiv \phi(z) $ satisfy the fourth-order ODE
\begin{equation}\label{eq3.1}
\phi_{zzzz} - q \phi_{zz} + p \phi = - \frac{1} {\beta} \left (
\phi^2 \right )_{zz},
\end{equation}
where \b z \equiv  x -ct, \qquad p \equiv  \frac {\gamma} {\beta},
\qquad q \equiv  - \frac {c} {\beta}.  \label{eq3.2}\e

\n Recently, it was proved by Liu and Varlamov \cite{LiVa} that
solitary waves $ \phi$ exist if the speed $ c $ satisfies that $ c <
2 \sqrt {\gamma \beta}, \;  \beta > 0 $ with the zero mass, that is,
$\int \phi \text{d}z= 0. $ Surprisingly, even if the mass of the
solitary wave of the KdV equation is not zero,  it is shown
\cite{LeLi} that the limit of the solitary waves of the Ostrovsky
equation tends to the solitary wave of the KdV equation as the
rotation parameter $ \gamma $  tends to zero. For $ \beta < 0, $
solitary waves in the form of stationary localized pulses cannot
exist at all \cite{GaSt91, Le81, VaLi04}.

\vskip 0.15cm

Equation (\ref{eq3.1}) is invariant under the transformation $ z \to
- z $ and thus it is a reversible system. In this section, we use
the theory of reversible systems
\cite{IoKi92,Ki88,Lo96,Lo97,Ha93,Ha97} to characterize the
homoclinic orbits to the fixed point of (\ref{eq3.1}), which
correspond to pulses or solitary waves of the Ostrovsky equation in
various regions of the $ (p, q) $ plane.

The linearized system corresponding to (\ref{eq3.1})
\begin{equation}\label{eq3.3}
 \phi_{zzzz} - q \phi_{zz} + p \phi = 0
\end{equation}
has a fixed point
\begin{equation}\label{eq3.4}
 \phi = \phi_z = \phi_{zz} = \phi_{zzz} = 0.
\end{equation}
Solutions $ \phi = k e^{\lambda z} $ satisfy the characteristic
equation
\begin{equation}\label{eq3.5}
 \lambda^4 - q \lambda^2 + p = 0,
\end{equation}
from which one may deduce that the structure of the eigenvalues is
distinct in four different regions of the $ (p, q) $ plane. These
regions, as well as the eigenvalue structure, are shown in Fig.1.
The regions are labeled as (1) -- (4). The boundaries of these
regions are the curves marked $ C_0 $ to $ C_3 $ in Fig.1. In
delineating the structure of the homoclinic orbits to fixed point
(\ref{eq3.4}) in various parts of the $ (p, q)$ space, we shall
first consider the bounding curves $ C_0 - C_3 $ and their
neighborhoods. Following this, we shall discuss the possible
occurrence and multiplicities of homoclinic orbits to (\ref{eq3.4}),
corresponding to pulse solitary waves of the Ostrovsky equation in
each of regions (1) through (4):

\noindent {\bf a.    Near $ C_0$:}  \ This curve, on which the
eigenvalues have the structure $ \lambda_{1-4} = 0, \ 0, \ \pm
\lambda, \  \lambda > 0 $ and its vicinity have been considered in
the context of reversible systems in \cite{IoKi92, Ki88}. In this
region, a standard analysis yields the normal form on the center
manifold \b \dot X_1 &=& X_2, \nonumber\\
\dot X_2 &=& \text{sign} (\mu) X_1 -  \frac {3}{2} X_1^2,\nonumber
\e where $ \mu $ is an unfolding parameter \cite{IoKi92}. For $ \mu
> 0, $ this yields a unique symmetric homoclinic solution
\b X_1(t) = \text{sech}^2 \left ( \frac {t} {2} \right ) \nonumber
\e in the vicinity of $ C_0$. One may also show persistence of this
homoclinic solution in the original system (\ref{eq3.4}) for $ \mu >
0 $ \cite{IoKi92}.

\vskip 0.2cm

\n {\bf b.  Near $ C_1$:} \ Near $ C_1, $ which corresponds to the
eigenvalue structure $ \lambda_{1-4} = 0, \ 0, \ \pm i \omega, \
\omega > 0, $ analysis of a four-dimensional normal from
\cite{IoKi92} shows that on the side of $ C_1 $ corresponding to
Region 3 in Fig.1 there is a $\text{sech}^2$ homoclinic orbit.
However, in Region 3, where the eigenvalue structure is that of a
saddle-center $ \lambda_{1-4} = \pm \lambda,  \ \pm i \omega, $ the
fixed point (\ref{eq3.4}) is non-hyperbolic. In fact it can be shown
by the ``antisoliton" method \cite{GaSt91} that there are no soliton
solutions of the system (\ref{eq3.1}) for which the function $
\phi(z) \to 0 $ when $ |z| \to \infty $ together with its
derivatives.

\vskip 0.2cm

\noindent {\bf c.    Near $ C_2$:}  \ In this region, where $
\lambda_{1-4} = \pm i \omega, \, \pm i \omega, $ analysis of
complicated normal form \cite{BlTiBrCoIo87, IoPe93} shows the
possible occurrence of so-called ``envelope" homoclinic solutions of
the form  $\text{sech}(kt) e^{i \alpha \theta}$ and with oscillating
tails in the so-called ``subcritical'' form. However, occurrence or
persistence of these solutions in the full nonlinear system
(\ref{eq3.1}) is a non-trivial issue and each  system must be
analyzed separately \cite{BuGr96, IoKi90, IoPe93}.

\vskip 0.2cm

\noindent {\bf d.  Near $ C_3$:}  \ There is no small-amplitude
bifurcation on $C_3,$ on which  $ \lambda_{1-4} = \pm  \lambda,  \
\pm \lambda   $ and the fixed point (\ref{eq3.4})  remains
hyperbolic. However, as we discuss below, there is a bifurcation
across it causing the creation of an infinite multiplicity of
homoclinic orbits.

\vskip 0.2cm

We turn next to each of the regions 1 to 4 in Fig.1 to discuss the
possible occurrence and multiplicity of homoclinic orbits in each
region.

\vskip 0.2cm

\noindent {\bf Region 1.}  \ In this region $ \lambda_{1-4} = \pm
\lambda,\  \pm i \omega $ and the fixed point (\ref{eq3.4}) is a
saddle focus.  Using a Shil'nikov type analysis, one may show
\cite{Ha93, Ha97} for general reversible systems such as (\ref
{eq3.1}) that the existence of one symmetric homoclinic orbit
implies the existence of an infinity of others. Hence,  we expect
our system (\ref {eq3.1}) to admit an infinity of such symmetric
$N$-pulses for each $ N > 1. $ Here, a symmetric $N$-pulse
oscillates $ N$ times in phase-space for $ z \in ( -\infty, \,
\infty) $ (or, more technically, crosses a transverse section to the
primary symmetric $ 1 $ pulse $ N $ times). In the context of the
Ostrovsky equation, these would be $ N$-peaked solitary waves, and
we expect an infinite family for all $ N > 1 $  for parameters $ (p,
q) $ in Region $ 1 $ of Fig.1.

\vskip 0.2cm

\noindent {\bf  Region 2.}  \ In this region $ \lambda_{1-4} = \pm
\lambda_1, \ \pm \lambda_2 $ and the fixed point (\ref{eq3.4}) is a
hyperbolic saddle point. Thus, there is no a priori reason for
multiplicity of homoclinic orbits in this region. However, depending
on the actual form of the nonlinear term, a symmetric homoclinic
orbit to (\ref{eq3.4}) may exist (see \cite{BuChTo96, ChTo93}).
Also, depending on further conditions \cite{ AlGrJoSa97, SaJoAl97},
a further ``orbit-flip" bifurcation may cause complex dynamics in
its neighborhood. In the context of our system (\ref{eq3.1}), these
issues will need further investigation to establish possible
existence of solitary wave solutions in this region of $ (p, q) $
space.

\vskip 0.2cm

\noindent {\bf Region 3.} \ The generic situation in this region has
already been considered in the discussion above pertaining to the
region near curve $ C_1. $

\vskip 0.2cm

\noindent {\bf  Region 4.} \ In this region, $ \lambda_{1-4} = \pm i
\omega, \ \pm i \omega_2 $ and (\ref{eq3.4}) is a focus. No
homoclinic orbits are known to exist in general here, although
complex dynamics may occur \cite{ArSe86, Se86}.

\section*{Acknowledgments}

R. I. Ivanov and Y. Liu gratefully acknowledge the hospitality and
support of the Mittag-Leffler Institute, Stockholm, where this
research was performed during the semester program on ``Wave Motion"
in the Fall of 2005.  R. I. Ivanov also acknowledges the funding
from the Irish Research Council for Science, Engineering and
Technology.

\begin{center}
\includegraphics[width=15cm]{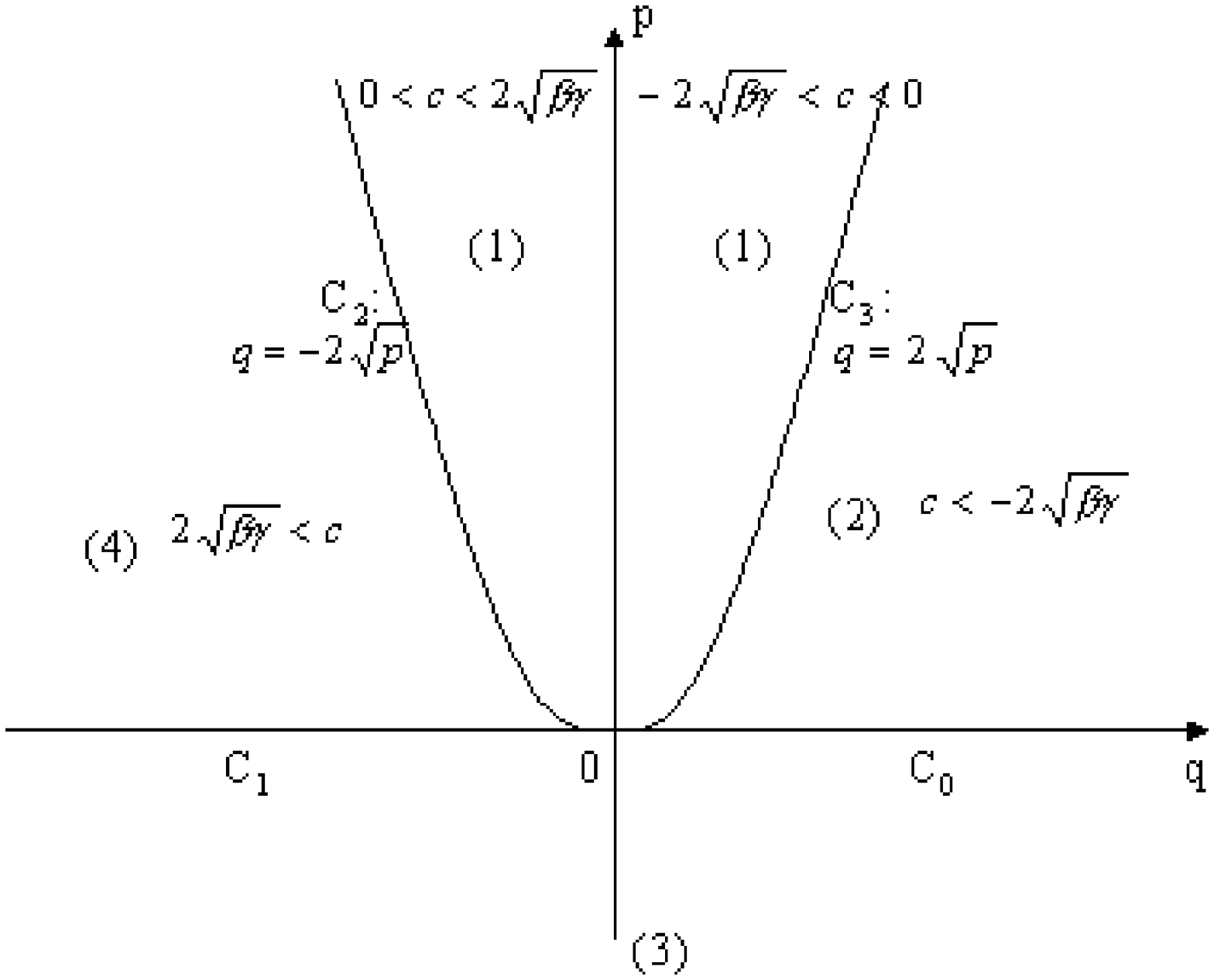}
 \centering {\Large Figure 1}
\end{center}





\end{document}